# Machine Learning Models for Dengue Forecasting in Singapore

Zi Iun Lai[1], Waikit Fung[2], Enquan Chew[1]

[1]Department of Mechanical Engineering, Stanford University

[2]Yong Loo Lin School of Medicine, National University of Singapore and National University Health System

## Abstract

With emerging prevalence beyond traditionally endemic regions, the global burden of dengue disease is forecasted to be one of the fastest growing. With limited direct treatment or vaccination currently available, prevention through vector control is widely believed to be the most effective form of managing outbreaks. This study examines traditional state space models (moving average, autoregressive, ARIMA, SARIMA), supervised learning techniques (XGBoost, SVM, KNN) and deep networks (LSTM, CNN, ConvLSTM) for forecasting weekly dengue cases in Singapore. Meteorological data and search engine trends were included as features for ML techniques. Forecasts using CNNs yielded lowest RMSE in weekly cases in 2019.

## Introduction

Dengue is a fast growing neglected tropical disease transmitted through the bite of mosquitos [1], [2]. Due in large part to global warming and increased connectivity, the incidence of dengue has increased 30-fold over the last half century [3], [4]. Estimates for yearly dengue prevalence in the 2010s range from 50-400 million cases, which was responsible for 1.14 disability-adjusted life years [5], [6], [7]. Approximately 50% of the world's population is at risk, with the geographical coverage of endemic regions [8], and resulting impact on human and economic health [9], expected to expand significantly. The existence of multiple dengue serotypes limits vaccine efficacy [10], and no specific treatment exists – vector control is widely recognized as the most effective measure for limiting the threat of dengue [11].

The city state of Singapore is dengue endemic with recurrent outbreaks and will be the focus of this study. Currently, the National Environmental Agency monitors potential dengue clusters by placing cylindrical traps known as Gravitraps in urban areas, and manually analyzing mosquitos caught for presence of virus [12], [13]. Detection of dengue in healthcare settings are also reported to the health ministry, and statistics on weekly cases are available [14]. However, these methods are retrospective; ideally, forecasts of dengue incidence can be obtained to perform vector control methods in advance.

## Dataset and Features

This project utilizes weekly dengue incidence in Singapore from 01-01-2012 to 11-29-2019, made available by the National Environment Agency for a total of 412 datapoints [14]. Past historical counts are naturally used for time series prediction. Vector dynamics is posited to have strong environmental dependence, as the primary vector, Aedes aegypti mosquitos, required stagnant water bodies to reproduce. The use of meteorological data for dengue forecasting have been explored by numerous studies [15], [16], [17], [18]; here, average daily total rainfall, mean temperature, wind speed, number of hours of sunshine, minimum and mean relative humidity are used as features. Weekly historical rainfall, temperature and windspeed data is available on the website of Meteorological Service Singapore [19], while aggregated monthly sunshine and relative humidity data was manually extracted from open source data by the Department of Statistics [20].

Inspired by reference [21], popularity of search terms 'mosquito', 'dengue symptoms' on Google were included as features. Weekly trends for search terms from 2012 to 2019 can be obtained by querying multiple shorter time frames on Google Trends [22]. Relative popularity of a search term in a given week is represented as a fraction compared to the largest count within the time period; concatenating trends can be performed by finding the relative peaks in each time frame followed by rescaling. Occurrence of spurious spikes, which could be due to panic – induced searching, were checked for, as in reference [23]. Sine and cosine of date-time were also used as features to capture potential seasonal patterns not reflected in meteorological features [24]. Missing values were replaced with local average. Plots of features against time are shown in Figure 1.

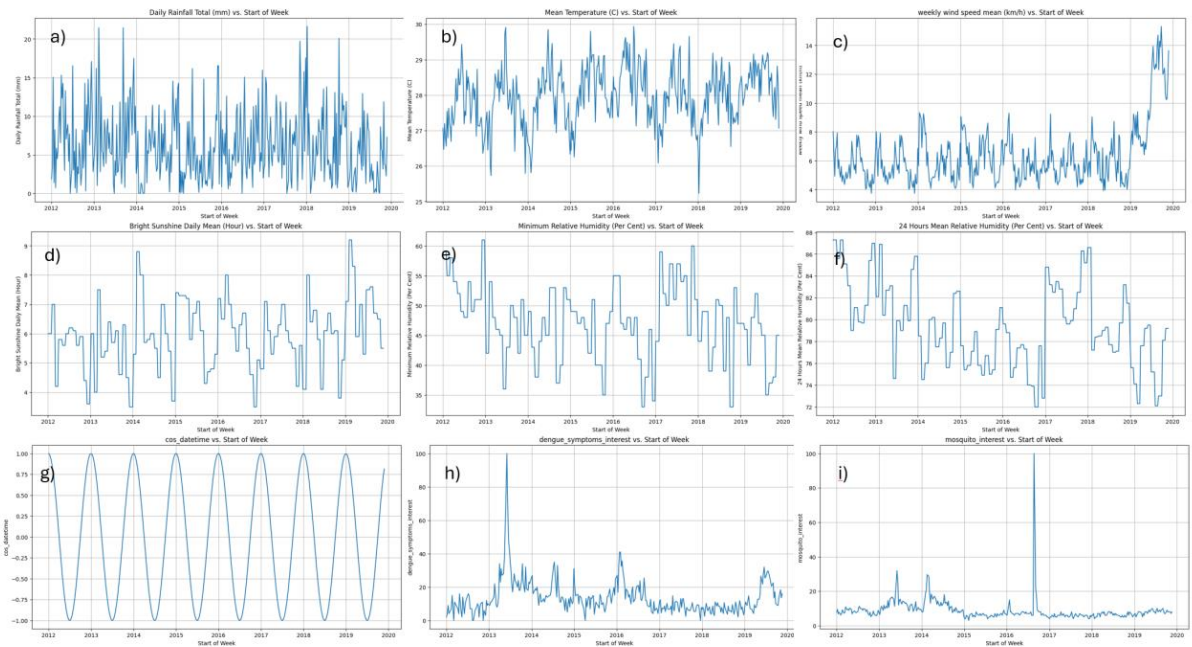

*Figure 1 – Plot of features against time. Weekly meteorological data: a) Daily total rainfall, b) Mean temperature, c) Mean wind speed. Monthly meteorological data: d) mean sunshine duration, e) minimum relative humidity, f) mean relative humidity. Seasonality: g) cosine datetime and sine datetime (not shown). Search term trends: h) for 'dengue symptoms', i) for 'mosquito'.*

Weekly dengue incidence labels were split into training, validation, and test sets, of sizes 311, 50 and 50 respectively. A plot of label against time is shown in Figure 2 – it should be noted that magnitude of patterns in the validation set differ from test set; nonetheless, models can be tuned using the validation set and a final evaluation of performance can be done using test set.

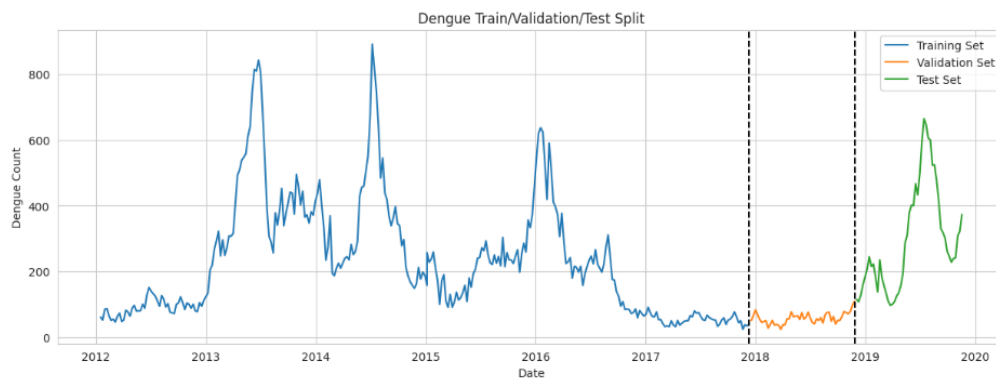

*Figure 2 – Weekly incidence of dengue in Singapore from 01-01-2012 to 29-11-2019, partitioned into training, validation and test sets.*

Pairwise correlations between features and weekly dengue incidences are shown in the figure below. Meteorological data was found to have weak correlation with case counts, while search terms displayed moderate to strong correlation. However, this does not guarantee search trends are good predictors of incidences, as media reports of dengue outbreaks could be responsible for spikes in search interests.

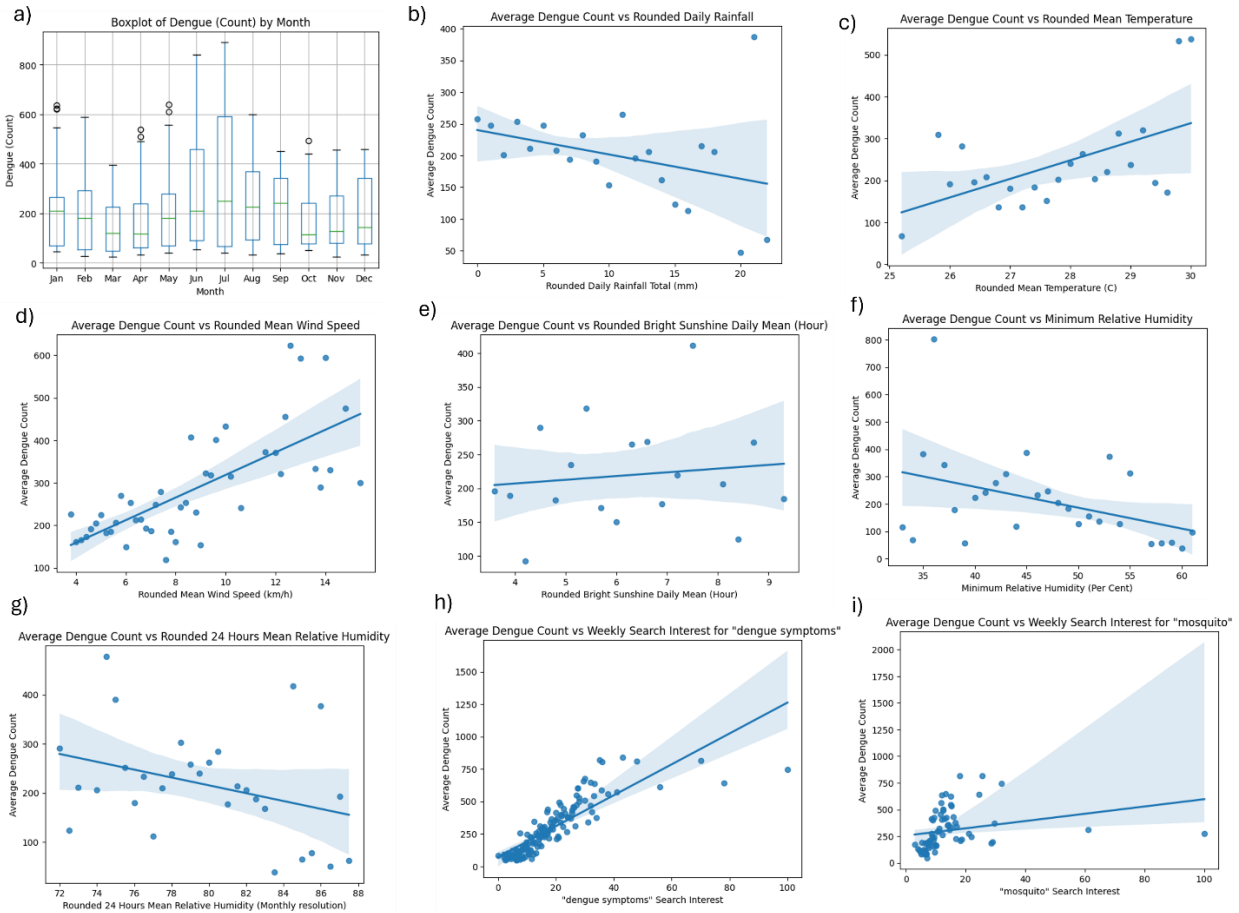

*Figure 3 – a): Boxplot of dengue incidences by months. Month as a categorical value accounted for 8.16% of variability, using one-way ANOVA tests. b): Scatter of dengue cases against average daily rainfall rounded to nearest integer. Person coefficient: -0.110. c): Scatter of dengue cases against mean temperature rounded to nearest 0.2C. Pearson coefficient: 0.210. d): Scatter of dengue cases against mean wind speed, rounded to nearest 0.2km/h. Pearson coefficient: 0.268. e): Scatter of dengue case counts against mean duration of bright sunshine, rounded to nearest 0.3h. Pearson coefficient: 0.083. f): Scatter of case counts against minimum relative humidity. Pearson coefficient: -0.194. g): Scatter of case counts against mean relative humidity, rounded to nearest 0.5%. Pearson coefficient: -0.232. h): Scatter of case counts against relative search interest for "dengue symptoms". Pearson coefficient: 0.794. i): Scatter of case counts against relative search interest for "mosquito". Pearson coefficient: 0.337. Scatter plots are shown with best fit line and 95% CI bands.*

## Baseline State Space Models

State space models work directly on time series data to uncover stochastic processes. Errors are assumed to be identically and independently distributed to estimate parameters using maximum likelihood estimation. Such models assume linear correlation structures in data [25]. In a moving average model, dengue incidences in a given week is forecasted by taking the average of past $x$ weeks' count. Because no weights are learnt, its performance can be examined across the entire dataset. Interestingly, optimal $x$ was found to be one, with RMSE of 44.611 and 13.439 on the whole and validation sets respectively.

An autoregressive model considering the past two weeks' data was fitted on the data. Achieving an RMSE of 14.654 on the validation set, the least squares linear regression equation was found to be:

$$\hat{x}_t = 1.024 x_{t-1} - 0.062 x_{t-2} + 8.841$$

Non-stationary characteristics, such as shifts in mean, variance, or autoregressive structure, can be captured by incorporating differencing. For the Autoregressive Integrated Moving Average (ARIMA) model, two lag observations were again used for autoregressive component, one difference was applied, and two lag errors were used to correct moving average component. One-step rolling forecast resulted in validation RMSE of 14.417. Finally, seasonality is encoded using a Seasonal ARIMA (SARIMA) model. Validation RMSE of 13.771 was achieved by including a seasonal period of 52 weeks and one seasonal lag observation. ARIMA and SARIMA models are established statistical methods in epidemiology, and were widely employed for dengue forecasting before the advent of machine learning techniques [26], [27], [28], [29]. Figure 4d illustrates forecasts of the above models.

## Supervised Learning Methods

Subsequent methods explore utilization of meteorological and search query data. Additional seasonality awareness was represented using the month as a feature. To leverage historical data, the previous two weeks' dengue counts were appended as features (referred to as 'lag1' and 'lag2' features).

### XGBoost

XGBoost is a popular tree ensemble technique which uses gradient boosting to minimize loss at each stage [30]. Grid search was used to identify the following optimal hyperparameter values: learning rate = 0.01, maximum depth = 3, number of trees = 500, which after fitting on training data, led to validation RMSE of 12.340. Figure 4a, e displays the relative importance of features used in the regressor as well as plot of predicted values against date.

## Support Vector Machine

Several reports have explored SVM outbreak prediction [31] and case forecasting [32], [33]. Typically used for classification, SVM uses kernel functions to cast data into higher dimensions in hopes of finding linearly separable hyperplanes. In regression, SVM instead identifies hyperplanes which fits the most points [34]. Predictions for query points are made based on their distances to hyperplanes. Linear kernel with regularization parameter C = 0.05 was found to be optimal. Suggested feature importance, as indicated by coefficient magnitude, and performance on validation data are shown in Figure 4e. SVM regression attained a validation RMSE of 13.649.

## K-Nearest Neighbors

Given a query point, KNN regressors identify top k nearest neighbors in feature space, then return the mean count as prediction. Using the optimal k = 3, a validation RMSE of 14.109 was achieved. Relative feature importance can be inferred from Shapley additive explanation values – lag values, minimum humidity and rainfall had larger importance (Figure 4c).

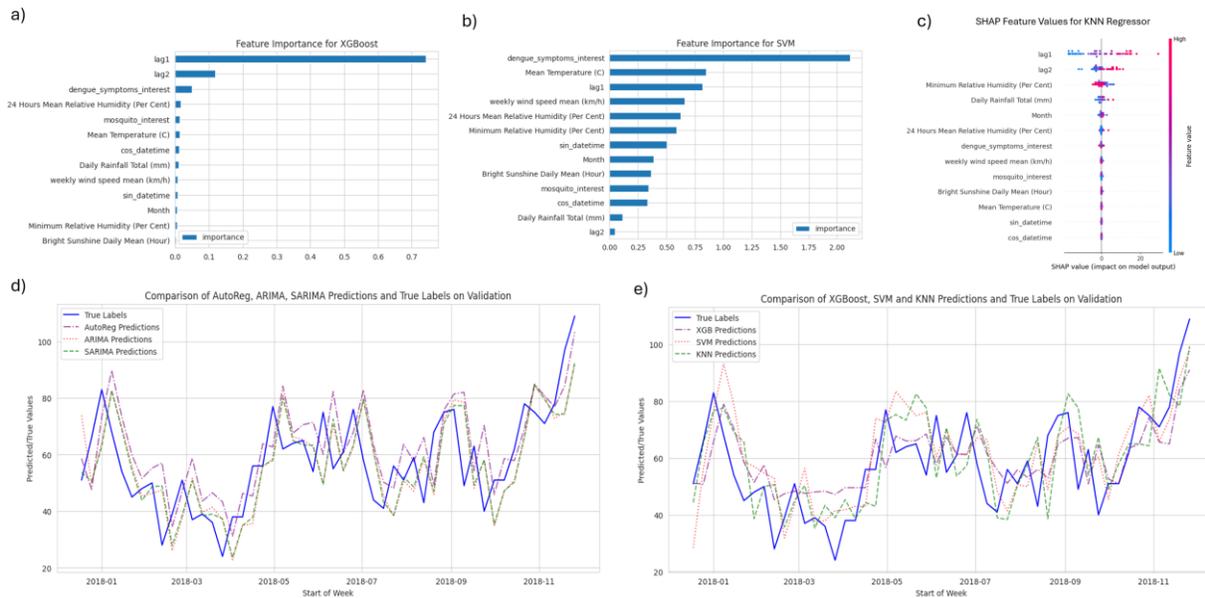

*Figure 4 – a) Relative importance of feature weights in XGBoost regressor, b) Relative feature importance for SVM regressor, c) SHAP feature scores for KNN regressor, d) Plot of predictions from state space models on validation set, e) Plot of supervised learning models on validation set.*

## Deep Learning Methods

### Long Short-Term Memory (LSTM)

A variant of Recurrent Neural Networks, LSTMs capture long term dependencies in time series data and address the vanishing gradient problem by using gates to regulate the flow during backpropagation [35]. Numerous studies have investigated LSTM for multivariate spatial-temporal dengue forecasting across different territories and reported impressive results [36], [37], [38, p. 20], [39]. The data is first reshaped into the following: number of samples, window size, number of features. Optimal window size was determined to be two.

The use of bidirectional LSTMs was inspired by reference [39], and architecture is similarly replicated with three bidirectional layers, one dropout of 0.2 and dense layer for regression. Unlike the report, grid searching on this dataset suggested the use of rectified linear activation function instead of hyperbolic tangent, with 128 hidden nodes per layer and learning rate of 0.001. The final dense layer predicted a continuous value with linear activation function. An RMSE of 13.040 was attained.

### Convolutional Neural Networks (CNN)

1-dimensional CNNs utilizes convolutional layers to extract temporal patterns for time series prediction. While less frequently used for time series prediction, it has been argued that the efficacy of CNNs in capturing local patterns leads to better forecasting on trends with localized temporal dependencies [40]. While less popularly used for time series prediction, CNNs were used to predict dengue counts using climate data [41]. The following model and hyperparameters were chosen through grid search: two 1D convolutional layers with kernel size = 2 followed by two dropout layer – hidden dense layer pairs. Predicted value was returned through a final dense layer. Rectified linear activation function was used, with optimal nodes per layer = 256, learning rate = 0.001, dropout ratio = 0.2. Validation RMSE of 12.841 was attained.

### Hybrid Networks (ConvLSTM)

Convolutional and LSTM layers can be concatenated to form a hybrid ConvLSTM architecture, which have been used in Covid19 and neurodegenerative disease predictions [42], [43]. It is hypothesized that through first identifying important local signatures using convolution, LSTMs can then better use them to capture temporal patterns. The architecture is adopted from reference [44] with minor tuning: three convolutional layers with kernel size = 3, one max-pooling with pool size = 2, three bidirectional LSTM layers with dropout of 0.25. All layers contained 128 nodes and used ReLU activation, except for the final dense layer. Validation RMSE of 13.01 was obtained.

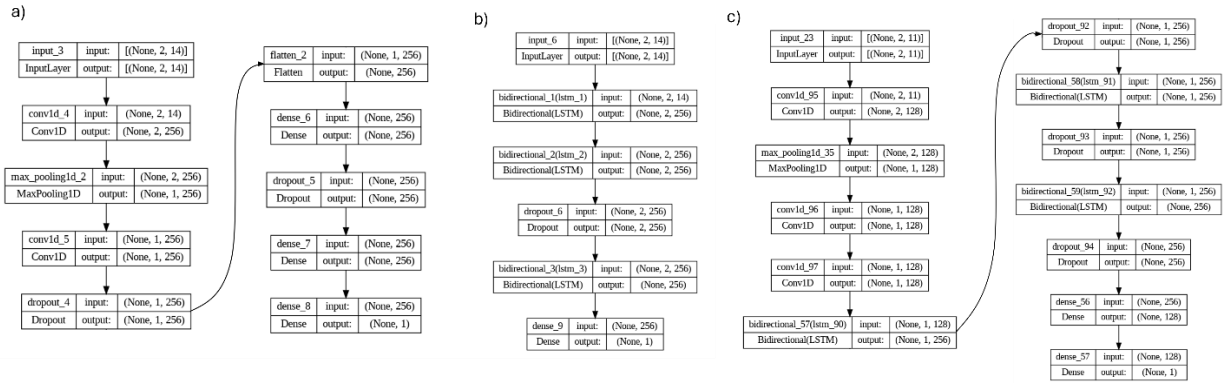

*Figure 5 – Model architectures for a): LSTM, b): CNN, c) hybrid ConvLSTM networks used.*

## Evaluation and Discussion

As the best performing supervised learning method, deep learning method and state space method respectively, XGBoost, CNN and SARIMA models were evaluated on test set (Figure 6a). CNN model yielded the lowest RMSE of 39.831. A comparison of model performance on validation and test sets is shown in Figure 6b.

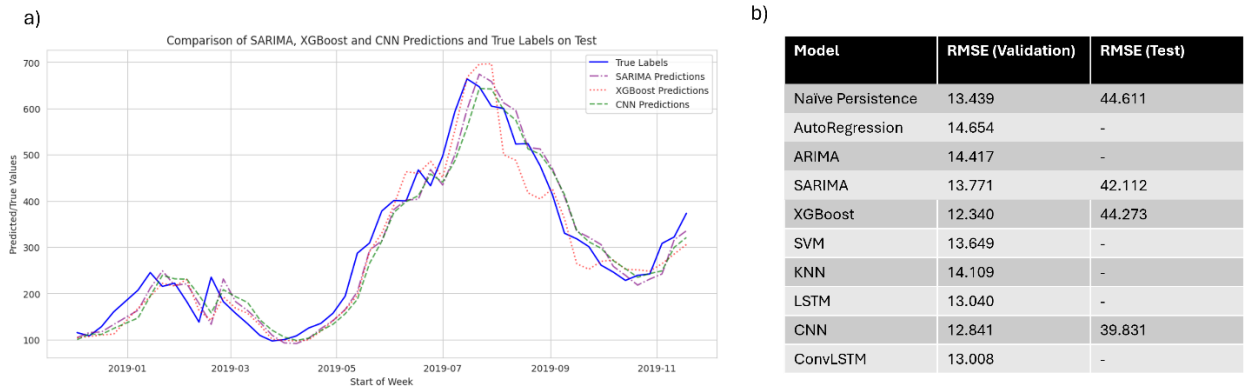

*Figure 6 – a) Comparison of SARIMA, XGBoost and CNN on Test set, b) table of RMSE of models on validation and test sets.*

However, several limitations in this study should be acknowledged. Firstly, the good performance of a trivial one-step naïve forecasting indicates that dengue counts exhibit, at best, a moderate dependence on the exogeneous features examined. The significant variation in data, such as that seen between validation and test dataset, requires further consideration of local features and refinement of vector-agent modelling. The rather small dataset (412 points across 8 years) necessitates validation on dengue trends in other regions. Despite that, it is interesting that leveraging exogenous meteorological and search trend features allowed deep models to outperform traditional state space techniques.

## Conclusion and Future Work

This report compared classical state space models, supervised learning, and deep learning methods for dengue prediction in Singapore. Results suggest utilizing meteorological and search trend data could aid in dengue forecasting. Moving forward, other public health models and neural network architectures, such as deep belief networks and transformers, can be explored. If local case data is available, spatial-temporal forecasting could be explored. Additionally, modelling transmission dynamics could allow targeted and preventative measures for vector control.